\documentclass{article}
\usepackage{spconf,amsmath,graphicx}
\usepackage{times}
\usepackage{epsfig}
\usepackage{amssymb}
\usepackage{xcolor}
\usepackage{url}
\usepackage[linesnumbered,ruled,vlined]{algorithm2e}
\usepackage[noend]{algpseudocode}
\usepackage{comment}
\usepackage{caption}
\usepackage{multirow}
\definecolor{vyellow}{rgb}{0.7490,0.5647,0}
\definecolor{vyelloworig}{rgb}{1,0.7529,0}
\definecolor{vmagenta}{rgb}{0.4392,0.1882,0.6274}
\definecolor{vmagenta3}{rgb}{0.7254,0.6352,0.7960}
\definecolor{vmagentaorig}{rgb}{0.9215,0.8039,1}
\definecolor{vblue}{rgb}{0.3686,0.7921,0.8431}
\definecolor{vred}{rgb}{0.7529,0,0}

\newcommand\beforecaptions{\vspace{-2mm}}
\newcommand\aftercaptions{\vspace{-4mm}}

\newcommand{\net}{SVAE-SR\@\xspace}

\renewcommand{\paragraph}[1]{\vspace{1mm} \noindent \textbf{#1}}

\title{SOFT-INTROVAE FOR CONTINUOUS LATENT SPACE IMAGE SUPER-RESOLUTION}
\name{Zhi-Song Liu\thanks{corresponding email: zhisong.liu@dell.com}, Zijia Wang, Zhen Jia}
\address{Dell Research}
\begin{document}

\maketitle

\begin{abstract}
Continuous image super-resolution (SR) recently receives a lot of attention from researchers, for its practical and flexible image scaling for various displays. Local implicit image representation is one of the methods that can map the coordinates and 2D features for latent space interpolation. Inspired by Variational AutoEncoder, we propose a Soft-introVAE for continuous latent space image super-resolution (\net). A novel latent space adversarial training is achieved for photo-realistic image restoration. To further improve the quality, a positional encoding scheme is used to extend the original pixel coordinates by aggregating frequency information over the pixel areas. We show the effectiveness of the proposed SVAE-SR through quantitative and qualitative comparisons, and further, illustrate its generalization in denoising and real-image super-resolution.
\end{abstract}
\section{Introduction}
\label{sec:intro}

Image super-resolution (SR) aims to enlarge the low-resolution (LR) images to the larger desired high-resolution (HR) images. It is widely used in digital display, broadcast and data compression/restoration. With various display devices and data resolutions, a flexible arbitrary image SR model can adjust different needs to produce image/videos with best visual experiences. Most existing state-of-the-art SR methods~\cite{ABPN,EDSR,RCAN,RDN,ESRGAN,DSRVAE} either focus on fixed super-resolution solutions (one model for one dedicated upsampling scenario), or integral upsampling scales (2$\times$, 4$\times$ or 8$\times$). They result in costly training efforts and imperfect image resolutions.

Instead, continuous image super-resolution~\cite{LIIF,METASR} provides arbitrary image/video scaling with photo-realistic visual quality. The goal is to discover the hidden latent space where the missing pixels can be estimated by the continuous feature representation. The advantage is that it can adjust size-varied display devices and reduce many training efforts when applying out-of-distribution super-resolution tasks. 

However, continuous image super-resolution tends to generate over-smooth images and is sensitive to noise. In order to produce clean photo-realistic super-resolution images, we propose Soft-introVAE for continuous latent space image super-resolution (SVAE-SR). The novelty is to use an autoencoder to discover the continuous image distribution space, where we condition the continuous LR features for supervision. The reused encoder works adversarially against the decoder for discriminating between real and generated samples. A soft threshold function is utilized to replace the hard margin in the evidence lower bound (ELBO). Furthermore, the positional encoding is embedded as the frequency expansion to introduce more pixels for prediction. To sum up, our key claims are 1) Soft-introVAE for arbitrary image super-resolution that can measure the adversarial conditional distribution of SR and HR images for reconstruction, and 2) a positional encoding scheme is modified to involve more neighborhood pixels for estimation.

\section{RELATED WORKS}
\label{sec:format}

\noindent \textbf{Implicit neural representation.} The idea of implicit neural representation is to use multi-layer perceptron (MLP) to learn pixels or other signals from coordinates. It has been widely used in 3D shape modeling~\cite{shape_1,shape_2}, surface reconstruction~\cite{surface_1,surface_2}, novel view rendering~\cite{NERF,structure_1,structure_2,nerf_2} and so on. For instance, Mildenhall et al.~\cite{NERF} propose to map the camera poses to the pixel values via a multi-layer MLP network. They use multiple-view images to optimize the network for implicit feature representation. Instead of using a voxel or point cloud, the implicit neural representation can 1) capture the fine details of scenes for photo-realistic reconstruction, and 2) also emit complex 3D representation as a small number of differentiable network parameters.

\begin{figure*}[ht!]
\vspace{-5mm}
	\begin{center}
		\centerline{\includegraphics[width=0.98\textwidth]{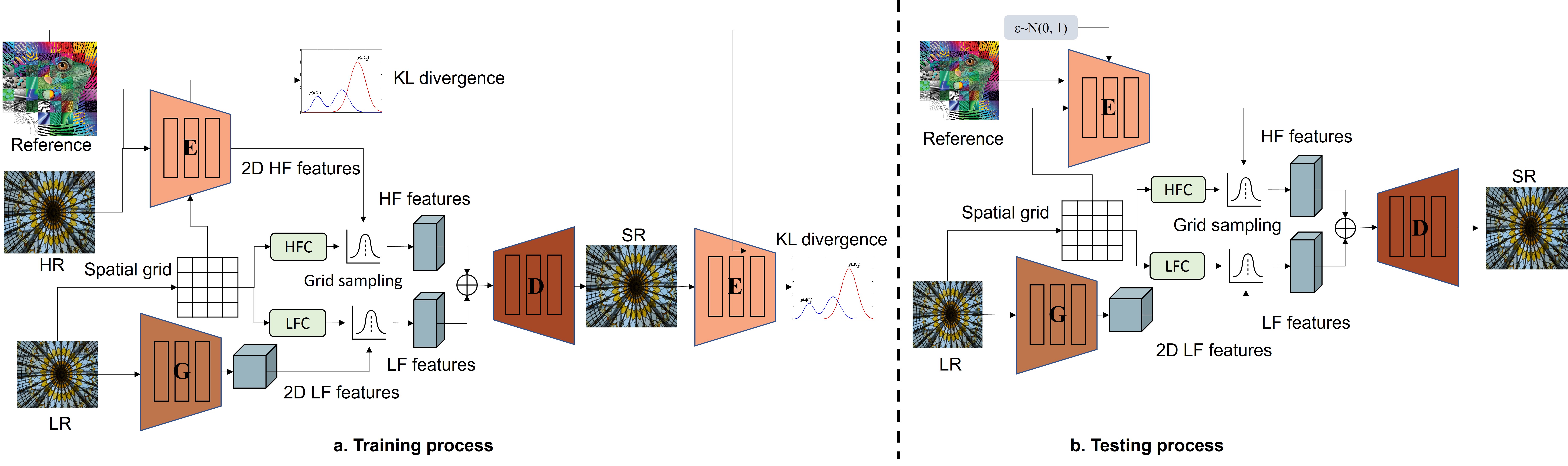}}
		\beforecaptions
		\caption{\small{\net. (a) At training, it uses an encoder (E), a decoder (D) and the feature extractor (G) to learn the continuous latent space distribution. Given the reference image as the conditional feature, the encoder works as a discriminator to identify whether the SR distribution is close to the HR distribution. The decoder learns the local implicit image function to have arbitrary scales of upsampling.
		(b) At test time, \net samples a random vector in the trained latent space and combines it with the reference features for high-frequency signal estimation. A final LIIF is trained to map the continuous features to the estimated RGB pixels.  
		}}
		\aftercaptions
		\label{fig:Figure1}
	\end{center}
	\aftercaptions
	%\vspace{-4mm}
\end{figure*}

\noindent \textbf{Image super-resolution.} The goal of image super-resolution is to enlarge LR images to the desirable resolutions with high visual quality. Depending on the metrics of evaluation, it can categorize into distortion-based SR~\cite{ABPN,EDSR,RCAN,RDN,HAN,SAN} and perception-based SR~\cite{ESRGAN,DSRVAE,REFVAE}. For the former one, using deeper neural networks with sophisticated designs usually leads to lower pixel distortions. For example, EDSR~\cite{EDSR} proposes a network using more convolutional kernels and layers for optimization. RDN~\cite{RDN} proposes a residual dense network to allow feature sharing. Most recently, attention~\cite{Attention} is also used in image SR~\cite{HAN,SAN} to involve more neighborhood pixels for estimation. For perception-based SR, GAN~\cite{GAN} and VAE~\cite{VAE} are two major architectures used for photo-realistic reconstruction. ESRGAN~\cite{ESRGAN} proposes a patch based GAN network to supervise the perceptual quality via a binary discriminator. SRVAE~\cite{DSRVAE} achieves the goal using a conditional VAE to minimize the distribution divergence. However, all the methods are for fixed upsampling factors, like 2$\times$ and 4$\times$. To have an arbitrary super-resolution solution, LIIF~\cite{LIIF} proposes to use the Local Implicit Image Function (LIIF) to explore continuous feature representation for SR. MetaSR~\cite{METASR} uses a Meta-Upscale Module to weight the upsampling filters for prediction dynamically.

\section{METHOD}
\label{sec:pagestyle}
Here, we describe the proposed Soft-Intro Variational AutoEncoder for continuous latent space image Super-Resolution (\net). \net performs continuous super-resolution by projecting image features into the latent space, and it conditions the LR features with random coordinate space sampling and transfers the new features to the decoder for reconstruction. The encoder is reused as a discriminator to distinguish the distributions of HR (real) and SR (fake) images.

\paragraph{Overview.} 
To learn the ground truth HR images $\mathbf{Y}$, \net is a feed-forward network (shown in Figure~\ref{fig:Figure1}) that reconstructs SR images $\mathbf{Y}'$ from LR images $\mathbf{X}$ and reference images $\mathbf{R}$. It consists of an encoder (E), a decoder (D) and a feature extractor (G). The encoder learns the latent vector of the joint HR-Reference distribution. The decoder uses the local implicit image function (LIIF) to transform the 2D features into pixel-coordinate pairs for 2D grid based interpolation. The proposed positional encoding scheme process the 2D grid by high-frequency coding (HFC) and low-frequency coding (LFC) to split the frequency bands. The LFC is with the LR features for low-frequency component reconstruction and the HFC is with the latent features for high-frequency reconstruction.

\paragraph{Soft-IntroVAE.} 
Overall, \net works as a soft-IntroVAE to explicitly extract the latent distribution for pixel reconstruction. Different from existing CNN/GAN based SR methods~\cite{ABPN,ESRGAN}, using VAE~\cite{DSRVAE} has been proven to be useful for robust photo-realistic quality restoration for real image SR. Soft-IntroVAE~\cite{soft-introvae} demonstrates its outstanding performance in image generation. By combining the advantages of VAE and GAN, it adversarially optimizes the encoder as an introspective discriminator. It fits the task of image SR that it can model the pixel correlations as multivariate Gaussian distributions $z\sim Q_{\omega}(z|\mathbf{X})=\mathit{N}(z;\mu_i,\sigma^2_i\mathit{I})$. We further modify it to a conditional Soft-IntroVAE model that it uses a reference image as the condition to learn the missing high-frequency signal. Formally, we give the mathematical description of the proposed conditional Soft-IntroVAE as,

\begin{footnotesize}
\vspace{-1mm}
\begin{equation} \tag*{(1)}
	\begin{matrix}
	\begin{split}
        & L_{E_\phi}(\mathbf{Y}|\mathbf{R},z)=ELBO(\mathbf{Y}|\mathbf{R})-\frac{1}{\alpha}exp\left(\alpha ELBO(D_\theta(z))\right) \\
        & L_{D_\theta}(\mathbf{Y}|\mathbf{R},z)=ELBO(\mathbf{Y}|\mathbf{R})+\gamma ELBO(D_\theta(z)) \\
        & where\quad ELBO(\mathbf{Y}|\mathbf{R}) = \mathbb{E}_{z\sim Q(z|\mathbf{Y},\mathbf{R})}[log P_\theta (\mathbf{Y},\mathbf{R}|z)] \\
        &\quad\quad\quad - D_{KL} [Q_\phi (z|\mathbf{Y},\mathbf{R})||P(z)]\leq log P_\theta(\mathbf{Y}|\mathbf{R})
	\label{eq:Equation1}
	\end{split}
	\end{matrix}
\end{equation}
\vspace{-1mm}
\end{footnotesize}

\noindent In Eq~\ref{eq:Equation1}, it can be seen that the process includes two steps: 1) fix the decoder and optimize the encoder to distinguish through the ELBO value, between HR images (high ELBO) and SR images (low ELBO) and 2) fix the encoder and optimize the decoder to ``fool'' the encoder with photo-realistic SR images. In such a way, we can push the distribution of the SR data close to the HR data. Note that the natural advantage of using Soft-IntroVAE is that we use a \textbf{soft exponential function over the ELBO} to improve the training stability. Meanwhile, using a reference image as the condition to learn the joint $p(\mathbf{Y}|\mathbf{R})$ probability so that the matched features from the reference images can be extracted for aiding reconstruction.

\begin{figure}[ht!]
\vspace{-5mm}
	\begin{center}
		\centerline{\includegraphics[width=\columnwidth]{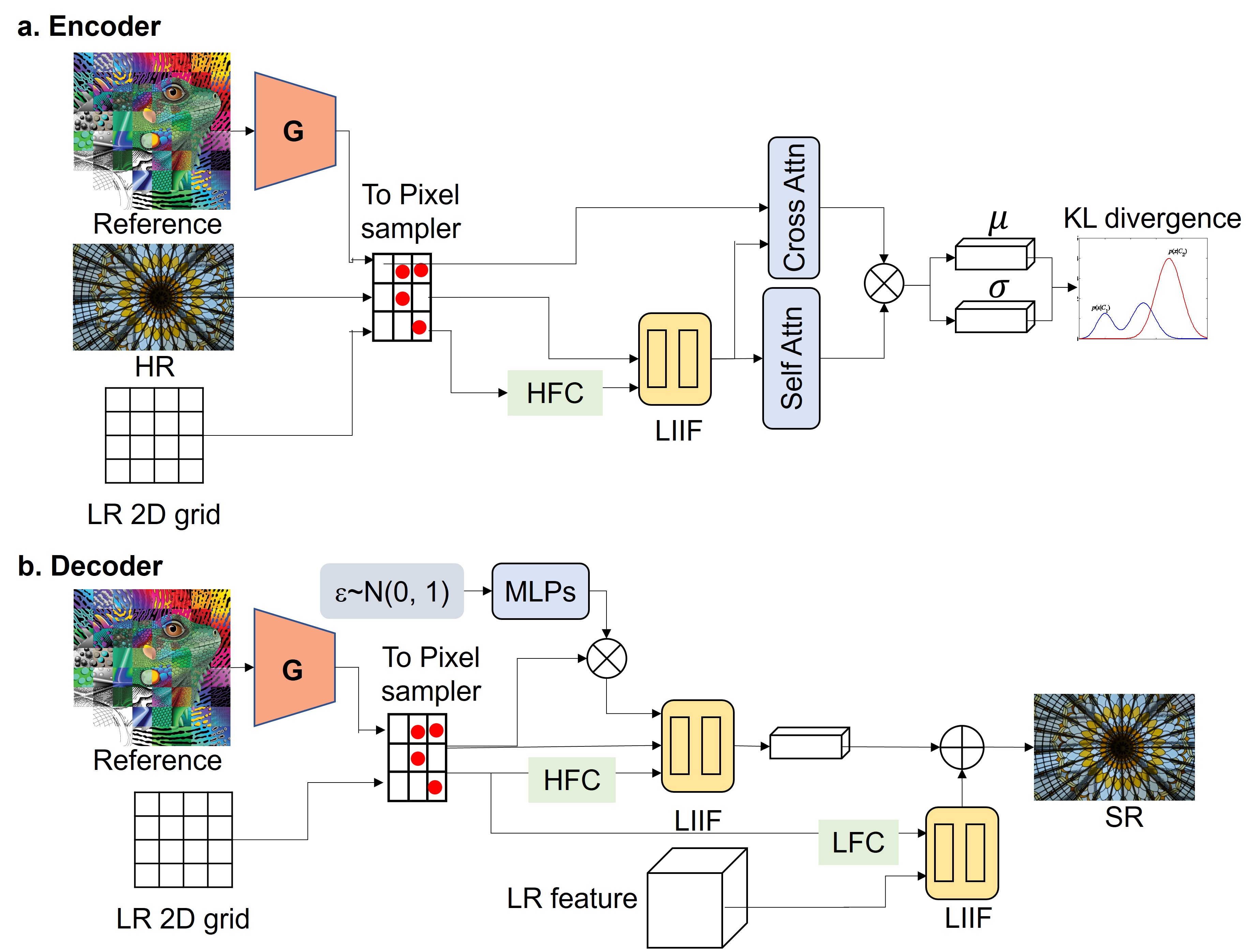}}
		\beforecaptions
		\caption{\small{The detailed structure of the encoder and decoder. The encoder takes the HR/SR image and reference image as inputs to computer their features correlations for joint distributions. The high-frequency positional Encoding (HFC) ensures the estimated features expand to the high-frequency bands. The decoder combines the LR features and sampled high-frequency features and pass them to the LIIF model for pixel estimation.
		}}
		\aftercaptions
		\label{fig:Figure2}
	\end{center}
	\aftercaptions
	%\vspace{-4mm}
\end{figure}

For clarification, the detailed structure of the encoder and decoder is shown in Figure~\ref{fig:Figure2}. For the encoder, it takes the reference, HR/SR image and 2D grid as inputs. A \textit{to pixel sampler} converts them to pixel-coordinate 1D vectors. The self-attention and cross-attention is applied to learn the latent distribution. The decoder combines the HF signal and LF signal together to form the final SR image.

\paragraph{Positional Encoding for frequency band splitting.} 
Another key component of the proposed \net is positional encoding (PE). Inspired by~\cite{NERF}, we expand the 2D grid map to much wider frequency bands. It can better fit the data with a high-frequency variation. It is specifically useful for image SR because the missing details around the edges and textures are high-frequency signals. Neural networks are biased to low-frequency reconstruction, which leads to over-smooth visual quality. Using PE can explicitly lift the feature for high-frequency mapping. Mathematically, given signal $o$,

\begin{footnotesize}
\vspace{-1mm}
\begin{equation} \tag*{(2)}
\lambda(o)=\left( sin(2^0\pi o), cos(2^0\pi o),..., sin(2^{L-1}\pi o), cos(2^{L-1}\pi o) \right)
\label{eq:Equation2}
\end{equation}
\vspace{-1mm}
\end{footnotesize}

\noindent Eq~\ref{eq:Equation2} maps the signal to an L-degree frequency band. We further split it into two parts, the first $L//2$ frequency bands for the low-frequency coding (LFC) and the rest for the high-frequency coding (HFC).

\paragraph{Training Loss. }
We train \net using the $l_1$ loss between SR and HR images and KL divergence as follows: 

\begin{footnotesize}
\vspace{-1mm}
\begin{equation} \tag*{(3)} \mathit{L}=\mathit{L}_{1}+\lambda ||Y'-Y||^1+\beta KL[Q_{\phi}(z|\mathbf{Y}|\mathbf{R})||\mathit{N}(0,1)],
\label{eq:Equation3}
\end{equation}
 \vspace{-1mm}
\end{footnotesize}

\noindent where $\lambda$ and $\beta$ are the weighting parameters to balance pixel distortion and KL losses. 

\section{EXPERIMENTS}
\label{sec:typestyle}

\paragraph{Implementation details.} 
We train our model on DIV2K~\cite{DIV2K} and Flickr2K~\cite{EDSR} datasets. They both contain images with resolutions larger than 1000$\times$1000. Same as~\cite{LIIF}, we first extract LR patches with the size of 48$\times$48. Then we randomly sample upsampling scales $\alpha$ in uniform distribution $U(1,4)$. The corresponding HR patches 48$\alpha \times$48$\alpha$. are then used to sample $48^2$ pixels to form coordinate-RGB pairs. For the reference image, we randomly choose one from Wikiart~\cite{WikiArt} that is widely used in style transfer. The testing datasets include Set5~\cite{Set5}, Set14~\cite{Set14}, Urban100~\cite{Urban100} and DIV2K validation~\cite{DIV2K}. We train with Adam optimizer with a learning rate of $10^{-4}$ and a batch size of 32 for 100k iterations (8hrs on two NVIDIA V100 GPUs). 

\begin{table}[t]
\caption{PSNR comparison between ours and other state-of-the-art methods in various upsampling scales. RDN trains different models for different scales. MetaSR, LIIF, and ours use one model for all scales and are trained with continuous random scales uniformly sampled in $\times1\sim\times$4. We also test on out-of-distribution scenarios in $\times$6 and $\times$8.}
\label{tab:sota}
\vskip -0.1in
\vspace{-2mm}
\begin{center}
\begin{small}
\resizebox{\linewidth}{!}{
\begin{tabular}{c|c|ccccc}
\hline
\multirow{2}{*}{Dataset} & \multirow{2}{*}{Method} & \multicolumn{3}{c}{In-distribution} & \multicolumn{2}{c}{Out-of-distribution} \\
 &  & $\times$2 & $\times$3 & $\times$4 & $\times$6 & $\times$8 \\ \hline
\multirow{4}{*}{Set5} & RDN & 38.24 & 34.71 & 32.47 & - & - \\
 & RDN-MetaSR & 38.22 & 34.63 & 32.38 & 29.04 & 26.96 \\
 & RDN-LIIF & 38.17 & 34.68 & 32.50 & 29.15 & 27.14 \\
 & \net(ours) & 38.23 & 34.72 & 32.60 & 29.23 & 27.24 \\ \hline
\multirow{4}{*}{Set14} & RDN & 34.01 & 30.57 & 28.81 & - & - \\
 & RDN-MetaSR & 33.98 & 30.54 & 28.78 & 26.51 & 24.97 \\
 & RDN-LIIF & 33.97 & 30.53 & 28.80 & 26.64 & 25.15 \\
 & \net(ours) & 34.01 & 30.58 & 28.86 & 26.72 & 25.23 \\ \hline
\multirow{4}{*}{B100} & RDN & 32.34 & 29.26 & 27.72 & - & - \\
 & RDN-MetaSR & 32.33 & 29.26 & 27.71 & 25.90 & 24.83 \\
 & RDN-LIIF & 32.32 & 29.26 & 27.74 & 25.98 & 24.91 \\
 & \net(ours) & 32.36 & 29.32 & 27.80 & 26.06 & 5.00 \\ \hline
\multirow{4}{*}{Urban100} & RDN & 32.89 & 28.80 & 26.61 & - & - \\
 & RDN-MetaSR & 32.92 & 28.82 & 26.55 & 23.99 & 22.59 \\
 & RDN-LIIF & 32.87 & 28.82 & 26.68 & 24.20 & 22.79 \\
 & \net(ours) & 32.92 & 28.88 & 26.73 & 24.29 & 22.87 \\ \hline
\end{tabular}
}
\end{small}
\end{center}
\vskip -3mm
\end{table}

\paragraph{General image super-resolution.} 
\net performs efficient super-resolution with high reconstruction quality. To show its effectiveness, we compare it to four state-of-the-art methods: Bicubic, EDSR~\cite{EDSR}, RDN~\cite{RDN}, LIIF~\cite{LIIF} and MetaSR~\cite{METASR} in Table~\ref{tab:sota}. For demonstration, we use continuous upsampling scales $\alpha$ in uniform distribution $U(1,4)$, and we also test on out-of-training-distribution scenarios, where larger unseen upsampling scales, namely 6$\times\sim 8\times$, are evaluated on unknown images. We can find that using our proposed method can achieve arbitrary image enlargement with superior performance across different datasets. Especially on out-of-distribution scenarios, ours performs even better with $+0.08\sim+0.09$ dB in PSNR. 

\begin{figure*}[ht!]
\vspace{-5mm}
	\begin{center}
		\centerline{\includegraphics[width=0.98\textwidth]{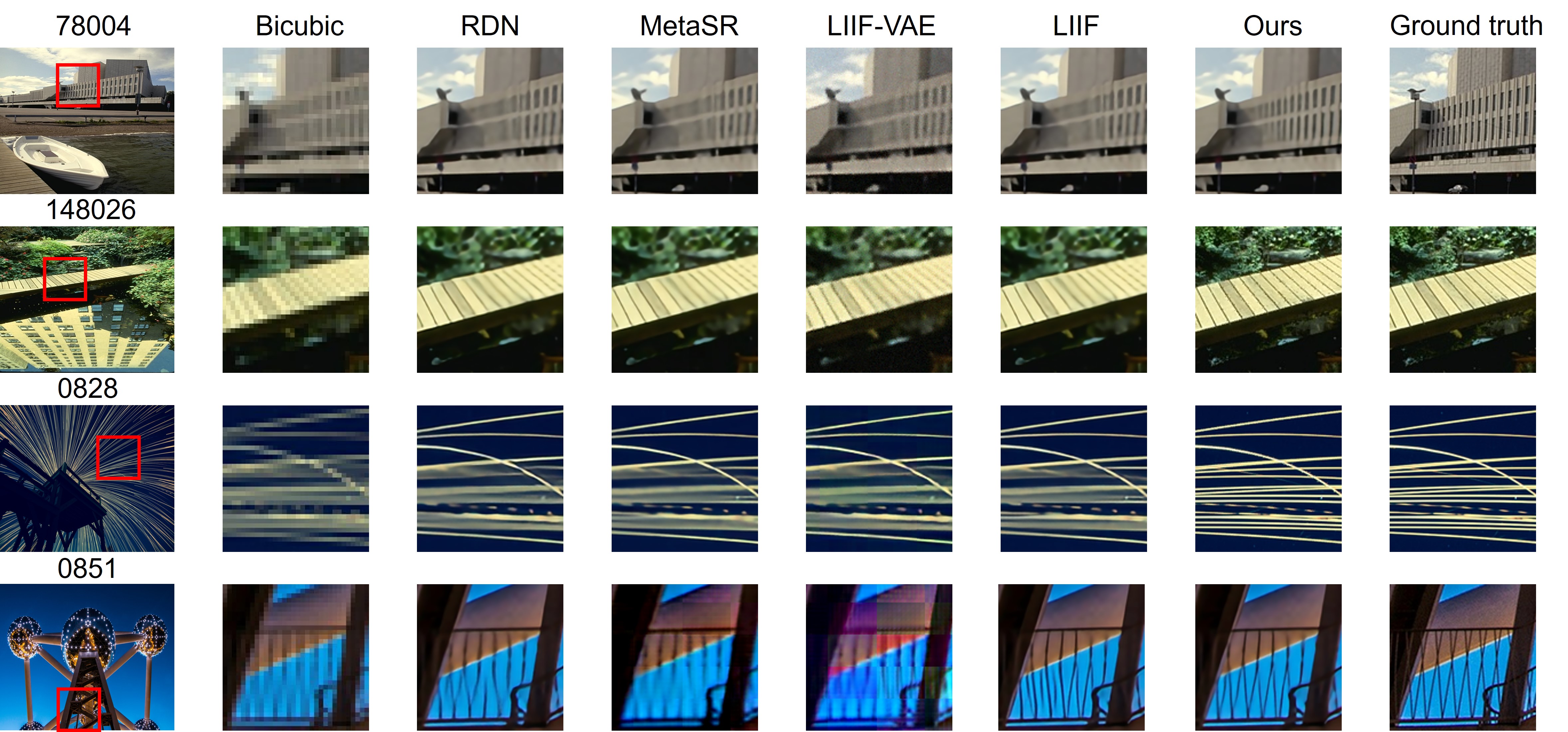}}
		\beforecaptions
		\caption{\small{Visual comparison to the state of the arts. We show 4$\times$ super-resolution on BSD100 (78004, 148026) and DIV2K-validation (0828,0851) datasets. We enlarge the regions in red boxes for better visualization.
		}}
		\aftercaptions
		\label{fig:Figure3}
	\end{center}
	\aftercaptions
	%\vspace{-4mm}
\end{figure*}

We show visual comparisons in Figure~\ref{fig:Figure3}. It can be seen that our method can restore the fine textures with better visual quality, like the windows in \textit{78004}, floor strides in \textit{148024}, lighting rays in \textit{0828} and the handrails in \textit{0851}.

\begin{table}[t]
\caption{{Abalation studies on our methods with or without Soft-IntroVAE and PE structures for image SR. The tests are done on DIV2K validation in PSNR(dB).}}
\label{tab:abalation}
\vskip -0.1in
\vspace{-2mm}
\begin{center}
\begin{small}
\resizebox{\linewidth}{!}{
\begin{tabular}{c|cccccc}
\hline
\multirow{2}{*}{Method} & \multicolumn{3}{c}{In-distribution} & \multicolumn{3}{c}{Out-of-distribution} \\
 & $\times$2 & $\times$3 & $\times$4 & $\times$6 & $\times$12 & $\times$18 \\ \hline
Bicubic & 31.01 & 28.22 & 26.66 & 24.82 & 22.27 & 21.00 \\
LIIF & 34.99 & 31.26 & 29.27 & 26.99 & 23.89 & 22.34 \\
LIIF-VAE & 34.89 & 31.20 & 29.16 & 26.90 & 23.81 & 22.29 \\
LIIF-GAN & 34.78 & 31.12 & 29.03 & 26.58 & 23.53 & 22.07 \\
LIIF-PE & 35.01 & 31.29 & 29.30 & 27.02 & 23.93 & 22.37 \\ \hline
\net w/o PE & 35.02 & 31.30 & 29.31 & 27.04 & 24.01 & 22.42 \\
\net w/ PE & 35.05 & 31.33 & 29.35 & 27.08 & 24.06 & 22.47 \\ \hline
\end{tabular}
}
\end{small}
\end{center}
\vskip -3mm
\end{table}

\paragraph{Ablation studies.}
\textbf{Effect of Soft-IntroVAE for SR.} To show the effectiveness of the proposed \net, we compare it with LIIF (baseline with pure MLP structure), LIIF-VAE (LIFF with ordinary VAE structure), LIIF-GAN (LIIF with ordinary GAN structure) and proposed \net. From 
 rows 2 to 4 and \net w/o PE in Table~\ref{tab:abalation}, we can see that with the same LIIF network, using Soft-IntroVAE (SVAE-SR w/o PE) achieve $+0.03\sim+0.12 dB$ in PSNR. On the other hand, using LIIF-GAN actually obtains lower PSNR, about $-2\sim-4 dB$. 

\begin{figure}[ht!]
\vspace{-2mm}
	\begin{center}
		\centerline{\includegraphics[width=\columnwidth]{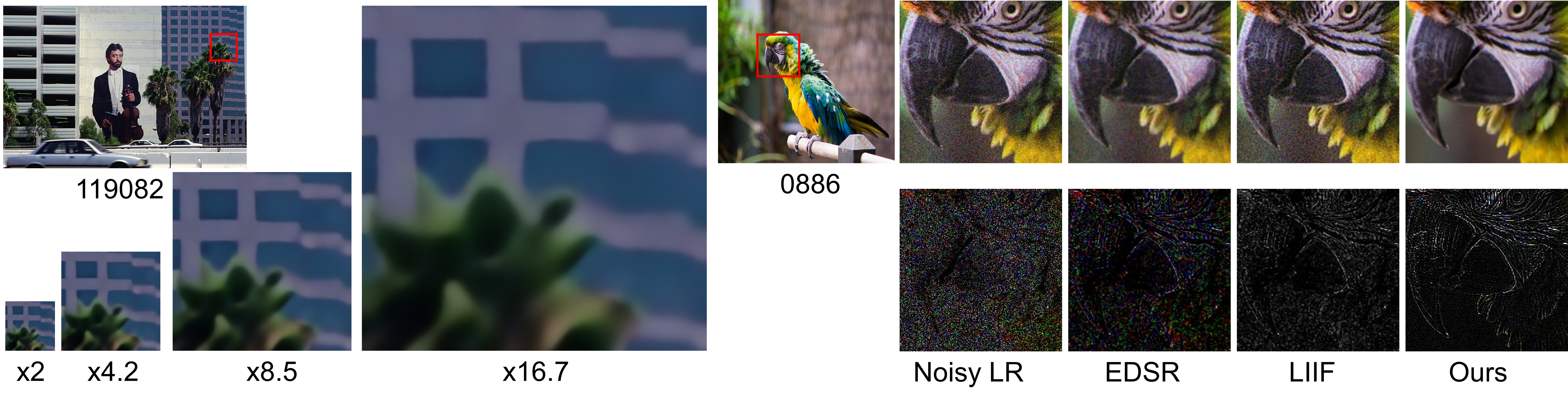}}
		\beforecaptions
		\caption{\small{Visualization on 1) $2\times\sim16\times$ continuous upsampling, and 2) image super-resolution on the noisy image with noise level $\tau=1.5$. We show the residual image between SR and HR images to enhance the differences.
		}}
		\aftercaptions
		\label{fig:Figure4}
	\end{center}
	\aftercaptions
	%\vspace{-4mm}
\end{figure}

\paragraph{Effect of positional encoding for SR.} To illustrate the effect of the positional encoding for SR. From Table~\ref{tab:abalation}, we conduct the comparisons among  LIIF (baseline with pure MLP structure) (row 2), LIIF-PE (baseline with positional encoding) (row 5) and ours with or without PE (row 6 and 7). We can see that using PE can improve the reconstruction quality by about $+0.03\sim0.05 dB$ in PSNR.

In Figure~\ref{fig:Figure4}, we show two more our results on continuous image super-resolution between $2\times\sim16\times$. We can see that the window patterns are well restored across different scenarios. We also mentioned that the proposed \net has the ability to overcome the noise for real image SR. We show one example using LR image with Gaussian random noise of intensity 1.5. We can see that our method achieves better results compared to other methods.

\paragraph{Computational cost.} 
Our method achieves relative real-time inference in a Nvidia V100 GPU, approximately 0.3s in 4$\times$ upsampling and 1.1s in 16$\times$ upsampling. 

\section{CONCLUSION}
In this paper, we propose a Soft-IntroVAE for continuous image super-resolution (\net). It combines the advantages of VAE and LIIF to explore the continuous latent space interpolation, which results in arbitrary upsampling with photo-realistic visual quality. In the meantime, the proposed positional encoding scheme expands the signal to much wider frequency bands, which avoids the network bias in the low-frequency domain. Experimental results on qualitative and quantitative comparisons show that our proposed \net achieves outstanding performance and points in a promising direction in robust real-image super-resolution.

% References should be produced using the bibtex program from suitable
% BiBTeX files (here: strings, refs, manuals). The IEEEbib.bst bibliography
% style file from IEEE produces unsorted bibliography list.
% -------------------------------------------------------------------------
{\small
\bibliographystyle{IEEEbib}
\bibliography{egbib}
}

\end{document}